\documentclass[%
reprint,
superscriptaddress,
nofootinbib,
amsmath,amssymb,
aps,
prd
]{revtex4-1}

\usepackage{physics}
\usepackage{graphicx}			
\usepackage{bm}				
\usepackage[caption=false]{subfig}
\usepackage{hyperref}			

\usepackage{soul}			
\usepackage[usenames]{color}		

\newcommand{\nn}{\nonumber}		

\allowdisplaybreaks

\begin{document}

\title{Massive Unruh particles cannot be directly observed}


\author{Filip Kia\l{}ka}
\email[]{filip.kialka@uni-duisburg-essen.de}
\affiliation{Institute of Theoretical Physics, University of Warsaw, Pasteura 5, 02-093 Warsaw, Poland}
\altaffiliation{Present address: University of Duisburg-Essen, Lotharstraße 1-21, 47057 Duisburg, Germany.}

\author{Alexander R. H. Smith}
\affiliation{Department of Physics and Astronomy, Dartmouth College, Hanover, New Hampshire 03755, USA}

\author{Mehdi Ahmadi}
\affiliation{Institute for Quantum Science and Technology, University of Calgary, Calgary, Alberta, Canada, T2N 1N4}
\affiliation{Department of Mathematics and Statistics, University of Calgary, Calgary, Alberta, Canada, T2N 1N4}

\author{Andrzej Dragan}
\affiliation{Institute of Theoretical Physics, University of Warsaw, Pasteura 5, 02-093 Warsaw, Poland}

\date{\today}	

\begin{abstract}
	We show that massive particles created in a relativistically accelerated reference frame, as predicted by the Unruh effect, can only be found in a tiny layer above the event horizon, whose thickness corresponds to a single Compton wavelength.
	This is beyond the reach of any detector and suggests that the Unruh effect may not ever be directly observed for massive fields. {The case of massless particles is also examined, for which qualitatively different behaviour is observed in a low-acceleration regime, suggesting that an observation of the Unruh effect for massless particles is more promising.}
\end{abstract}

\maketitle


\section{Introduction} 
\label{sec:intrduction}
	Quantum field theory on curved spacetimes examines the behaviour of quantum fields in situations where the effect of gravity may be treated classically and the back-reaction of the field on the spacetime geometry is ignored.
	Despite these restrictions, some of the most fascinating results of the last half-century in theoretical physics have emerged from this framework\,---\,most notably Hawking radiation, which predicts the creation of particles in the vicinity of black holes \cite{Hawking:1975}.
	This discovery has led to a deep connection between the laws of black hole mechanics and thermodynamics, offering a valuable hint as to what we should expect from a complete theory of quantum gravity \cite{Wald:1994}.

	The Unruh effect is an analogous phenomenon in a uniformly accelerated frame of reference: if the state of a quantum field is seen to be in the vacuum by an inertial observer, a uniformly accelerating observer will see themselves immersed in a thermal bath of particles at a temperature proportional to their proper acceleration \cite{Fulling:1973, Davies:1975, Unruh:1976}.
	Observing the Unruh effect poses a tremendous challenge, as the accelerations required to see a bath of particles at a temperature appreciably different from zero are of the order of $10^{19} \: \mathrm{g}$ (see, for example, Fig.~\ref{fig:part_vs_acc_massless} in Sec.~\ref{sec:number-massless} of this article).
	If we are ever to observe the Unruh effect directly, these particles will need to be detected.
	Our analyses suggest that such a direct observation for massive fields is unlikely, as these particles can only exist in a tiny layer (on the order of a Compton wavelength) above the Rindler horizon.

	One possible way of deriving the Unruh effect is to consider the expectation value of the particle number operator associated to an accelerating observer given that the state of the field is in the Minkowski vacuum \cite{Birrell:1982}.
	This number operator counts the total number of particles in the entire space as seen by an accelerating observer.
	However, any observation of the effect will necessarily involve a detector which is only able to count particles in a finite volume.
	Thus, the physically relevant question is not how many particles exist in the entire space for an accelerating observer, but how many particles will an accelerating observer detect in a finite volume? 
	
	Previous studies involved determining the quantum states of accelerating localized wave packets for both bosonic \cite{Dragan2013,Dragan2013B,Ahmadi2016} and fermionic \cite{Richter2017} fields.
	These studies indicate that the particle content of such individual wave packets is negligible.
	It is therefore important to ask about the total number of particles inside a finite volume, taking into account all orthonormal wave packets supported within that volume.

	In the field of quantum optics, the question of how many particles exist in a given state is well studied through the introduction of a number operator associated with a finite volume \cite{Mandel:1966,Mandel:1965,Mandel:1964,Jordan:1964}.
	We generalize this number operator so that it may be used for the purpose of counting particles in non-inertial frames.
	Noting that such a generalization is not unique, we consider two possible generalizations and show that the resulting number operators lead to the same conclusions in the case of the Unruh effect.
	
	This article is structured as follows.
	We begin in Sec.~\ref{sec:scalar_field} by reviewing the quantization of a real scalar field in 3+1D Minkowski and Rindler spacetimes.
	Then, in Sec.~\ref{sec:operators}, we introduce two finite volume number operators and discuss their properties.
	In Sec.~\ref{sec:number-massive} and Sec.~\ref{sec:number-massless}, we use these operators to compute the number of massive and massless particles seen by an accelerating observer within a finite volume and conclude with a discussion and outlook in Sec.~\ref{sec:discussion_outlook}.
	Throughout this article we use natural units $\mathrm{c} = \hbar =1$ and the metric signature~$(+\!~-\!~-\!~-)$.

\section{Scalar field in Minkowski and in Rindler coordinates} 
\label{sec:scalar_field}

	\begin{figure}[]
	   \centering
	   \includegraphics[width=3.2in]{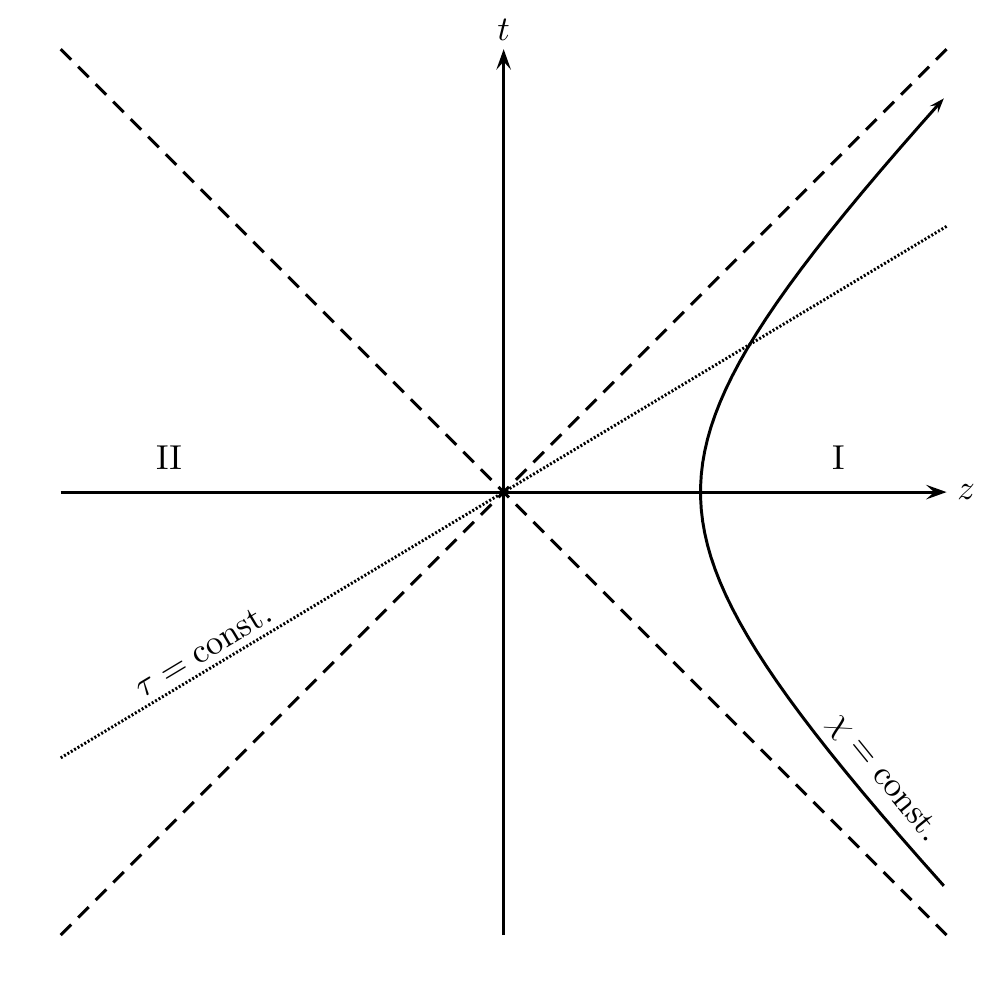}
	   \caption{
	   		Rindler coordinatization of Minkowski spacetime.
	   		The right Rindler wedge, labeled by I, is covered by the coordinates $(\tau, \chi , \bm{x}_\perp)$.
	   		The Rindler horizons are indicated by the dashed lines.
	   		}
	   \label{fig:Spacetime}
	\end{figure}

	Consider a real scalar field satisfying the Klein-Gordon equation
	\begin{align}
		\left(\Box + m^2 \right) \hat{\phi} = 0,
	\end{align}
	where $m$ is the mass of the field and $\Box\equiv \eta^{\mu \nu} \partial_\mu \partial_\nu$ is the d'Alembertian.
	In Minkowski coordinates, the normalized field eigenmodes are
	\begin{align} \label{eq:Minkowski_modes}
		u_{\bm{k}}(t, \bm{x} ) \equiv \frac{1}{\sqrt{2 \omega_{\bm{k}} (2 \pi)^3 }} e^{i \bm{k} \cdot \bm{x} - i \omega_{\bm{k}} t},
	\end{align}
	where $\bm{x}\equiv (x,y,z)$, $\bm{k}\equiv(k_x,k_y,k_z)$, and $\omega_{\bm{k}}\equiv~\sqrt{\bm{k}^2 + m^2}$.
	For brevity, the arguments of the mode functions will often be suppressed, $u_{\bm{k}} \equiv u_{\bm{k}}(t, \bm{x} )$.
	Modes~\eqref{eq:Minkowski_modes} form a complete orthogonal set,
	\begin{align}
		(u_{\bm{k}} ,u_{\bm{k'}}) = \delta^3(\bm{k}-\bm{k}'), \qquad
		(u_{\bm{k}}^* ,u_{\bm{k'}}) = 0,
	\end{align}
	with respect to the Klein-Gordon inner product.
	The latter is defined as
	\begin{align}
		\label{scalarproduct}
		(f , g) \equiv i \int_\Sigma \mathrm{d}^3x \,  f^*(t, \bm{x} ) \overset\leftrightarrow{\partial}_t g(t, \bm{x} ) ,
	\end{align}
	where $f$ and $g$ are functions on the hypersuface $\Sigma$, the integration is over the entire volume of the spatial slice $\Sigma$, and
	$f \overset\leftrightarrow{\partial}_t g \equiv f \partial_t g - g \partial_t f$.

	These Minkowski modes can be used to expand the field as
	\begin{align} \label{posnegdec}
		\hat{\phi}(t, \bm{x} ) &= \int \mathrm{d}^3k \, \left[u_{\bm{k}} \hat{a}_{\bm{k}} + u_{\bm{k}}^* \hat{a}_{\bm{k}}^\dagger \right] \nn \\
		&= \hat{\phi}^+(t, \bm{x} ) + \hat{\phi}^-(t, \bm{x} ),
	\end{align}
	where $\hat{a}_{\bm{k}}^\dagger$ and  $\hat{a}_{\bm{k}}$ are the creation and annihilation operators satisfying $[ \hat{a}_{\bm{k}}^\dagger, \hat{a}_{\bm{k}'}^\dagger] = [ \hat{a}_{\bm{k}}, \hat{a}_{\bm{k}'}] =0$, $[ \hat{a}_{\bm{k}}, \hat{a}_{\bm{k}'}^\dagger ] = \delta_{\bm{k}\bm{k}'}$, and where
	\begin{align}
	\label{posnegmink}
		\hat{\phi}^+(t, \bm{x} ) \equiv \int \mathrm{d}^3k \, u_{\bm{k}}  \hat{a}_{\bm{k}} \equiv \left[\hat{\phi}^-(t, \bm{x} ) \right]^\dagger.
	\end{align}
	The Minkowski vacuum state $\ket{0}_\mathrm{M}$ is defined as the state annihilated by all Minkowski annihilation operators
	\begin{align}
		\forall_{\bm{k}} \quad \hat{a}_{\bm{k}} \ket{0}_\mathrm{M} = 0.
	\end{align}

	The Rindler coordinates $(\tau, \chi , \bm{x}_\perp)$ (see Fig.~\ref{fig:Spacetime}) are defined by
	\begin{align} \label{eq:Rindler_trafo}
		t \equiv \chi \sinh a \tau, \quad \qty(x,y) \equiv \bm{x}_\perp, \quad z \equiv \chi \cosh a \tau,
	\end{align}
	where $a$ is a positive constant.
	Rindler coordinates are apt for describing accelerating observers, since their world lines correspond to lines of constant $\chi$.
	These trajectories are characterized by a proper acceleration of $1/\chi$ and are at a fixed distance $\chi$ from the Rindler horizon (dashed lines in Fig.~\ref{fig:Spacetime}).
	The temporal coordinate $\tau$ is also the proper time along the trajectory given by $\chi=1/a$.
	Coordinates given by Eq.~\eqref{eq:Rindler_trafo} cover only the right Rindler wedge, depicted as region I in Fig.~\ref{fig:Spacetime}; however, analogous coordinates may be introduced to cover the left Rindler wedge.

	The Klein-Gordon equation in Rindler coordinates takes the form
	\begin{equation}
		\qty(\frac1{(a \chi)^2} \frac{\partial^2}{\partial \tau^2} - \frac{\partial^2}{\partial \chi^2} - \frac1{\chi} \frac{\partial}{\partial \chi} - \frac{\partial^2}{\partial x^2} - \frac{\partial^2}{\partial y^2} + m^2) \hat{\phi} = 0.
	\end{equation}
	Its positive-frequency, normalized eigenmodes\footnote{
		Positive-frequency with respect to $\tau$ translations.
	} are:
	\begin{align} \label{defrindlermodes}
		w_{\Omega \bm{k}_\perp }(\tau,\chi, \bm{x}_\perp) &\equiv   \sqrt{\frac{ \sinh (\pi \Omega/a)}{4\pi^4 a}}  e^{i \bm{k}_\perp \cdot \bm{x}_\perp  - i \Omega \tau}\nonumber \\ 
		&\quad \times K_{i \Omega /a} \left(\sqrt{ \bm{k}_\perp^2 +m^2}\chi\right),
	\end{align}
	where $\bm{x}_\perp \equiv (x, y)$, $\bm{k}_\perp \equiv (k_x, k_y)$, and $K_\nu(x)$ is the modified Bessel function of the second kind~\cite{Crispino:2008a}.
	Again, we will often suppress the arguments of the mode functions, $w_{\Omega \bm{k}_\perp } \equiv w_{\Omega \bm{k}_\perp }(\tau,\chi, \bm{x}_\perp)$.
	Modes given by Eq.~\eqref{defrindlermodes} form a complete orthogonal set of solutions in the right wedge,
	\begin{align}
		(w_{\Omega \bm{k}_\perp } , w_{\Omega' \bm{k}_\perp' }) &= \delta(\Omega-\Omega')\delta^2( \bm{k}_\perp - \bm{k}_\perp'), \nn \\
		(w_{\Omega \bm{k}_\perp } , w^\star_{\Omega' \bm{k}_\perp' }) &= 0.
	\end{align}

	These Rindler modes can be used to expand the field as
	\begin{align}
		\hat{\phi}(\tau,\chi, \bm{x}_\perp) &= \int \mathrm{d}\Omega \mathrm{d}^2k_\perp \, \left[w_{\Omega \bm{k}_\perp } \hat{b}_{\Omega \bm{k}_\perp } +w^*_{\Omega \bm{k}_\perp } \hat{b}_{\Omega \bm{k}_\perp }^\dagger \right] \nn \\
		&=\hat{\phi}^+(\tau, \chi,\bm{x}_\perp ) + \hat{\phi}^-(\tau, \chi, \bm{x}_\perp ),
	\end{align}
	where $\hat{b}_{\Omega \bm{k}_\perp }^\dagger$ and  $\hat{b}_{\Omega \bm{k}_\perp }$ are creation and annihilation operators satisfying $[\hat{b}_{\Omega \bm{k}_\perp }, \hat{b}_{\Omega' \bm{k}_\perp '}] =0$, $[ \hat{b}_{\Omega \bm{k}_\perp }, \hat{b}_{\Omega' \bm{k}_\perp' }^\dagger ] = \delta(\Omega-\Omega')\delta^2( \bm{k}_\perp - \bm{k}_\perp')$, and
	\begin{align}
		\hat{\phi}^+(\tau,\chi, \bm{x}_\perp)&\equiv  \int \mathrm{d}\Omega \mathrm{d}^2k_\perp \, w_{\Omega \bm{k}_\perp } \hat{b}_{\Omega \bm{k}_\perp } \equiv \left[\hat{\phi}^-(\tau,\chi, \bm{x}_\perp) \right]^\dagger\label{psiI}.
	\end{align}
	The Rindler vacuum state $\ket{0}_\mathrm{R}$ is defined as the state annihilated by all Rindler annihilation operators
	\begin{align}
		\forall_{\Omega, \bm{k}_\perp} \quad \hat{b}_{\Omega \bm{k}_\perp } \ket{0}_\mathrm{R} = 0.
	\end{align}
	The Minkowski and Rindler vacuum states do not coincide, which is known as the Unruh effect~\cite{Unruh:1976,Fulling:1973}.
	As a consequence of this fact, an accelerating observer in the Minkowski vacuum experiences a thermal state of the field with a temperature proportional to their proper acceleration.
	In particular, for particles of a given frequency, the detection statistics is given by the Bose-Einstein distribution with the temperature independent of the frequency.

\section{Finite-volume particle number operators} 
\label{sec:operators}

	The number operator
	\begin{align}
		\hat{N} \equiv \int \mathrm{d}^3k \, \hat{a}_{\bm{k}}^\dagger \hat{a}_{\bm{k}} ,\label{totalNumberOperator}
	\end{align}
	counts the total number of particles seen by an inertial observer on an entire spatial hypersurface $\Sigma$.	However, any detector of finite size will only count particles within a volume $V < \Sigma$.
	Construction of an operator corresponding to a finite-sized particle-counting detector is well studied in quantum optics \cite{Mandel:1995,Mandel:1966,Mandel:1964}.
	There, one introduces the absorption operator
	\begin{align} \label{AbsorptionOperator}
			\hat{A}(t,\bm{x}) \equiv \int \mathrm{d}^3k \,\sqrt{2 \omega_{\bm{k}}} u_{\bm{k}} \hat{a}_{\bm{k}}.
	\end{align}
	Then, a number operator corresponding to a volume $V$, is defined as an integral of the particle density $\hat{A}^\dagger \hat{A}$
	\begin{align} \label{eq:Min_Man_op}
		\hat{N}_1(t,V) \equiv \int_V \mathrm{d}^3x \, \hat{A}^\dagger(t,\bm{x}) \hat{A}(t,\bm{x}).
	\end{align}
	When the volume considered is the entire hypersurface, $V=\Sigma$, the operator $\hat{N}_1(t,\Sigma)$ is equal to the total operator $\hat{N}$ in Eq.~\eqref{totalNumberOperator}.

	The above construction of the finite volume particle number operator is not unique (hence the subscript ``$1$'' in Eq.~\eqref{eq:Min_Man_op}).
	An alternative finite-volume particle number operator can be constructed by expressing the creation and annihilation operators as $\hat{a}_{\bm{k}}^\dagger = (\hat{\phi}^+, u_{\bm{k}})$, $\hat{a}_{\bm{k}} = (u_{\bm{k}},\hat{\phi}^+)$, with $\hat{\phi}^+$ given by Eq.~\eqref{posnegmink}, and consequently the number operator in Eq.~\eqref{totalNumberOperator} as
	\begin{align}
		\hat{N} &= \int \mathrm{d}^3k \, (\hat{\phi}^+, u_{\bm{k}}) (u_{\bm{k}},\hat{\phi}^+) = \left( \hat{\phi}^+, \hat{\phi}^+\right) \nn \\
		&= i \int_\Sigma \mathrm{d}^3x \, \hat{\phi}^-(t, \bm{x} ) \overset\leftrightarrow{\partial}_t \hat{\phi}^+(t, \bm{x} ) ,\label{Noperator}
	\end{align}
	where we have made use of the completeness of $u_{\bm{k}}$.
	Equation~\eqref{Noperator} expresses the total particle number operator as an integration of the density $i  \hat{\phi}^-(t, \bm{x} ) \overset\leftrightarrow{\partial}_t \hat{\phi}^+(t, \bm{x} )$ over $\Sigma$.
	This suggests a construction of an alternative finite volume number operator, defined as
	\begin{align} \label{OurNumb}
		\hat{N}_2(t,V) \equiv  i\int_V \mathrm{d}^3x\, \hat{\phi}^-(t, \bm{x}) \overset\leftrightarrow{\partial}_t \hat{\phi}^+(t, \bm{x}).
	\end{align}
	It is clear from the above definition that $\hat{N}_2(t,\Sigma) = \hat{N}$.

	The operator $\hat{N}_2(t,V)$ generalizes immediately to accelerating reference frames; one simply takes the $+/-$ split with respect to Rindler time rather than Minkowski time\footnote{
		The disadvantage of this construction is that, unlike operator~\eqref{eq:Min_Man_op}, \eqref{OurNumb} is not manifestly positive-definite.
		In fact, by evaluating the expectation value of~\eqref{OurNumb} on a state of the field containing two coherent excitations, one can show that the density $\big<i  \hat{\phi}^-(t, \bm{x} ) \overset\leftrightarrow{\partial}_t \hat{\phi}^+(t, \bm{x} )\big>$ can take negative values.
	}:
	\begin{align} \label{eq:ours_expanded}
		\hat{N}^\mathrm{R}_2(V,\tau) &\equiv i \int_V \frac{\mathrm{d}\chi}{a\chi} \mathrm{d}^2x_\perp \, \hat{\phi}^-(\tau,\chi, \bm{x}_\perp) \overset\leftrightarrow{\partial}_\tau \hat{\phi}^+(\tau,\chi, \bm{x}_\perp) \nonumber\\
		&=  \int_V  \frac{\mathrm{d}\chi}{a\chi}\mathrm{d}^2x_\perp \int \mathrm{d}\Omega \mathrm{d}^2 k_\perp \mathrm{d}\Omega' \mathrm{d}^2 k^\prime_\perp \,  (\Omega + \Omega') \nonumber\\
		&\quad \times w_{\Omega \bm{k}_\perp}w^*_{\Omega' \bm{k}^\prime_\perp} \hat{b}_{\Omega \bm{k}_\perp} \hat{b}^\dagger_{\Omega' \bm{k}^\prime_\perp}.
	\end{align}

	To generalize $\hat{N}_1(t,V)$ to accelerating reference frames, the Minkowski modes and ladder operators in Eq.~\eqref{AbsorptionOperator} must be replaced with their Rindler counterparts ($u_{\bm{k}} \mapsto w_{\Omega\bm{k}_\perp}$, $\omega_{\bm{k}} \mapsto \Omega$, $\hat{a}_{\bm{k}} \mapsto \hat{b}_{\Omega \bm{k}_\perp}$).
	This gives
	\begin{align} \label{eq:Rindler-abs-op}
			\hat{A}^\mathrm{R}(\tau,\chi,\bm{x}_\perp) \equiv \int \mathrm{d}\Omega\mathrm{d}^2k_\perp \,\sqrt{2 \Omega} w_{\Omega \bm{k}_\perp} \hat{b}_{\Omega\bm{k}_\perp}.
	\end{align}
	Substituting Eq.~\eqref{eq:Rindler-abs-op} into Eq.~\eqref{eq:Min_Man_op} yields
	\begin{align} \label{eq:Mandels_expanded}
		\hat{N}^\mathrm{R}_1 (V,\tau) &=\int_V \frac{\mathrm{d}\chi}{a\chi} \mathrm{d}^2x_\perp \int \mathrm{d}\Omega \mathrm{d}^2 k_\perp \mathrm{d}\Omega' \mathrm{d}^2 k^\prime_\perp \,  2\sqrt{\Omega\Omega'}  \nonumber\\
		&\quad \times w_{\Omega \bm{k}_\perp}w^*_{\Omega' \bm{k}^\prime_\perp} \hat{b}_{\Omega \bm{k}_\perp} \hat{b}^\dagger_{\Omega' \bm{k}^\prime_\perp}.
	\end{align}

	We see that the operators in Eqs.~\eqref{eq:ours_expanded} and~\eqref{eq:Mandels_expanded} differ by a factor $(\Omega' + \Omega)/(2\sqrt{\Omega' \Omega})$ within the integrand.
	As shown in the Appendix, this factor does not affect nonlocality\,---\,both operators do not commute with themselves for spacelike separated regions.
	Furthermore, the $(\Omega' + \Omega)/(2\sqrt{\Omega' \Omega})$ factor does not affect the expectation value for states $\ket{\psi}$ such that
	\begin{equation} \label{eq:delta_condition}
		\expval{b^\dagger_{\Omega\bm{k}_\perp} \hat{b}_{\Omega'\bm{k}'_\perp}}{\psi} \propto \delta( \Omega - \Omega' ).
	\end{equation}
	This condition is satisfied by the Minkowski vacuum $\ket{0}_\mathrm{M}$, as well as all other Fock states.
	This implies that the results that follow hold despite the ambiguity in defining finite-volume particle density operators \cite{Terno:2014, Celeri:2016}, at least when considering the two operators in Eqs.~\eqref{eq:ours_expanded} and~\eqref{eq:Mandels_expanded}.
	In what follows, we consider the field to be in the Minkowski vacuum state, and therefore we will write $\hat{N}^\mathrm{R}(V,\tau)$ making no distinctions between $\hat{N}^\mathrm{R}_1(V,\tau)$ and $\hat{N}^\mathrm{R}_2(V,\tau)$.
	We also drop the dependence on $\tau$, as the problem we consider is stationary due to the uniform acceleration of the observer.

\section{Number of massive Unruh particles in a finite volume} 
\label{sec:number-massive}

	\begin{figure}[t]
		\centering
		\includegraphics[width=0.48\textwidth]{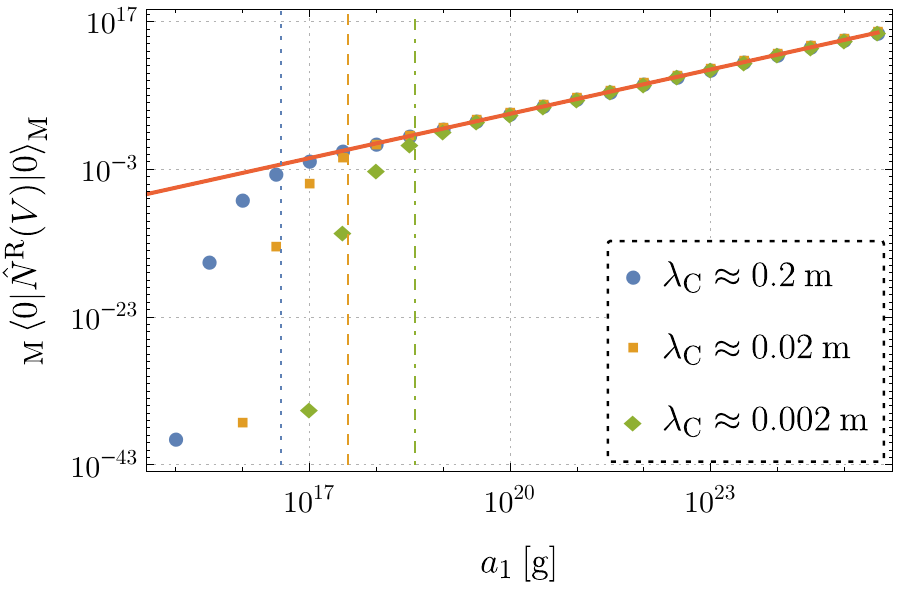}
		\caption{
		Average number of massive Unruh particles in a $1 \: \mathrm{m}^3$ box as a function of its proper acceleration.
		The field mass is $m=10^{-11}$ (dots), $m=10^{-10}$ (squares), and $m=10^{-9}$ (diamonds) electron masses, corresponding to Compton wavelengths shown in the inset.
		The number of particles does not depend on the length of the volume, $\chi_2 - \chi_1$, as long as it is greater than $\lambda_C$.
		The vertical lines indicate the accelerations for which the sudden birth of the Unruh effect occurs, as given by Eq.~\eqref{eq:sudden-birth}.
		The continuous line is the high-acceleration limit given by Eq.~\eqref{eq:jestem_wielki}.
		}
		\label{fig:part_vs_acc_massive}
	\end{figure}

	We now compute the expectation value $_\mathrm{M}\langle 0| \hat{N}^\mathrm{R}(V) | 0\rangle_\mathrm{M}$; i.e., the number of particles occurring due to the Unruh effect inside a finite volume $V$.
	This quantity involves a summation over all modes of the field and therefore can be considered an upper bound on the number of particles an accelerating, band-limited detector of volume $V$ would detect.

	We start with Eq.~\eqref{eq:ours_expanded} and use the relation \cite{Crispino:2008a}
	\begin{equation}
		_\mathrm{M}\langle 0| b^\dagger_{\Omega\bm{k}_\perp} \hat{b}_{\Omega'\bm{k}'_\perp}|0\rangle_\mathrm{M} =
		\frac{ \delta( \Omega - \Omega' ) \delta^2( \bm{k}_\perp - \bm{k}_\perp' ) }{\mathrm{e}^{2 \pi \Omega / a}-1}.
	\end{equation}
	If we assume that the region $V$ has a constant $x$-$y$ cross-section area $S_\perp$ and spans from $\chi_1$ to $\chi_2$ ($0<\chi_1<\chi_2$), we obtain
	\begin{equation}
		_\mathrm{M}\langle 0| \hat{N}^\mathrm{R}(V) | 0\rangle_\mathrm{M} = S_\perp \int \mathrm{d}\Omega \mathrm{d}^2 k_\perp \, 2\Omega \frac{ \int_{\chi_1}^{\chi_2} \frac{\mathrm{d}\chi}{a\chi} \, \abs{w_{\Omega \bm{k}_\perp}(\chi)}^2 }{\mathrm{e}^{2 \pi \Omega / a}-1}.
	\end{equation}
	Expanding $w_{\Omega \bm{k}_\perp}(\chi)$ using Eq.~\eqref{defrindlermodes}, changing the integration variables $\mathrm{d}^2k_\perp \mapsto k_\perp \mathrm{d}\varphi \mathrm{d}k_\perp$, and performing the integration over $\varphi$ leads to
	\begin{align} \label{eq:ours_expval_final}
		_\mathrm{M}\langle 0| \hat{N}^\mathrm{R}(V) | 0\rangle_\mathrm{M} &= \frac{S_\perp}{2\pi^3} \int\mathrm{d}\tilde{\Omega} \mathrm{d} k_\perp \, k_\perp \tilde{\Omega} \mathrm{e}^{- \pi \tilde{\Omega}} \nonumber \\
		 &\quad  \times \int_{\chi_1}^{\chi_2} \frac{\mathrm{d}\chi}{\chi} \,K^2_{\mathrm{i}\tilde{\Omega}}\qty(\sqrt{k_\perp^{2}+m^{2}}\chi),
	\end{align}
	where $\tilde{\Omega}\equiv\frac{\Omega}{a}$ is a dimensionless frequency.
	We note that the result does not depend on the choice of the parameter $a$ appearing in the Rindler coordinates defined in Eq.~\eqref{eq:Rindler_trafo}, as discussed in detail in~

	By evaluating Eq.~\eqref{eq:ours_expval_final} numerically, we can plot the average number of Unruh particles in a finite volume as a function of its proper acceleration (see Figs.~\ref{fig:part_vs_acc_massive} and~\ref{fig:part_vs_acc_massless}).
	We chose to parametrize the motion of the region with $a_1 \equiv 1/\chi_1$; i.e., the highest proper acceleration in the volume.

	The plot in Fig.~\ref{fig:part_vs_acc_massive} shows two unique properties of the Unruh effect for massive particles from the perspective of a finite-sized detector: the sudden birth of the effect for high acceleration (as indicated by the vertical lines) and the independence of the number of particles on their mass.
    However, the acceleration at which the Unruh effect appears is mass dependent.
    In the following subsections, we describe these effects in detail and demonstrate how they follow from Eq.~\eqref{eq:ours_expval_final}.

	\subsection{Sudden birth of the Unruh effect for massive particles\label{ssub:sudden_birth}} 

		To estimate the acceleration for which the sudden birth of the Unruh effect occurs, we use the approximation
		\begin{equation} \label{eq:bessel_approx}
		K_{\mathrm{i} \nu}\qty(x) \approx 0 \quad \text{for} \quad x > 1, \, \nu \leq 1.
		\end{equation}
		Contributions to the integral in Eq.~\eqref{eq:ours_expval_final} with $\nu > 1$ can be neglected because of the $\mathrm{e}^{-\pi \tilde{\Omega}}$ factor.
		The integral may thus reach non-negligible values only when the argument of the Bessel function becomes smaller than or comparable to unity.
		This leads to the conclusion
		\begin{align} \label{eq:sudden-birth}
		a_1\geq \frac{1}{\chi} \gtrsim \sqrt{k_\perp^{2}+m^{2}} \geq m.
		\end{align}
		Restoring the powers of $\hbar$ and $\mathrm{c}$ gives $a_1 \gtrsim m \mathrm{c}^3 / \hbar$, which corresponds to $\chi_1 \lesssim \hbar / m \mathrm{c}$; i.e., the distance between the volume $V$ and the Rindler horizon is smaller than the reduced Compton wavelength\footnote{
			A similar estimate is obtained by requiring that $k_\mathrm{B} T_\mathrm{U} \gtrsim m\mathrm{c}^2$, where $T_U$ is the Unruh temperature.
		}.
		Equation~\eqref{eq:sudden-birth} essentially means that the massive Unruh particles exist only in a Compton-wavelength-thick layer above the Rindler horizon.
		This is consistent with another observation we made by studying Eq.~\eqref{eq:ours_expval_final} numerically, namely that the number of Unruh particles is independent of the volume's length, $\chi_2 - \chi_1$, as long as it is greater than the Compton wavelength.

	\subsection{High-acceleration limit} 
	\label{ssub:massive_high_acceleration_limit}
		To demonstrate the mass-independence and the power-law visible in Fig.~\ref{fig:part_vs_acc_massive} for high accelerations, we change the integration variable in Eq.~\eqref{eq:ours_expval_final} from $\chi$ to $\chi' = \chi \sqrt{k_\perp^{2}+m^{2}}$.
		This gives
				\begin{widetext}
		\begin{align}
			_\mathrm{M}\langle 0| \hat{N}^\mathrm{R}(V) | 0\rangle_\mathrm{M} = & \frac{S_\perp}{2\pi^3} \int \mathrm{d}\tilde{\Omega} \mathrm{d} k_\perp \, k_\perp \tilde{\Omega} \mathrm{e}^{- \pi \tilde{\Omega}}  \int_{\chi_1 \sqrt{k_\perp^{2}+m^{2}}}^{\chi_2 \sqrt{k_\perp^{2}+m^{2}}} \frac{\mathrm{d}\chi'}{\chi'} \,K^2_{\mathrm{i}\tilde{\Omega}}\qty(\chi').
		\end{align}
		Furthermore, introducing $k' = \chi_1 k_\perp$ leads to

		\begin{equation} \label{eq:final_chi'_k'}
			_\mathrm{M}\langle 0| \hat{N}^\mathrm{R}(V) | 0\rangle_\mathrm{M} = \frac{S_\perp}{2 \pi^3 \chi_1^2} \int \mathrm{d}\tilde{\Omega} \mathrm{d} k' \, k' \tilde{\Omega} \mathrm{e}^{-\pi \tilde{\Omega}} \int_{\sqrt{(k')^2+(m \chi_1)^2}}^{\sqrt{(k' \chi_2 / \chi_1)^2+(m \chi_2)^2}} \frac{\mathrm{d} \chi'}{\chi'} \,K^2_{\mathrm{i}\tilde{\Omega}}\qty(\chi').
		\end{equation}
		The high-acceleration limit corresponds to $\chi_1 \to 0$, which gives
		\begin{equation} \label{eq:jestem_wielki}
			_\mathrm{M}\langle 0| \hat{N}^\mathrm{R}(V) | 0\rangle_\mathrm{M} \approx \frac{S_\perp}{2 \pi^3 \chi_1^2} \int \mathrm{d}\tilde{\Omega} \mathrm{d} k' \, k' \tilde{\Omega} \mathrm{e}^{-\pi \tilde{\Omega}} \int_{k'}^\infty \frac{\mathrm{d} \chi'}{\chi'} \,K^2_{\mathrm{i}\tilde{\Omega}}\qty(\chi') \equiv C S_\perp a_1^2.
		\end{equation}
		\end{widetext}
		We emphasize that Eq.~\eqref{eq:jestem_wielki} holds for any $m$ or $\chi_2$.
		The constant $C$ is independent of those parameters and is defined as
		\begin{align}
			C &\equiv \frac{1}{2 \pi^3} \int_0^\infty \mathrm{d}\tilde{\Omega} \, \mathrm{e}^{-\pi \tilde{\Omega}} \tilde{\Omega} \int_0^\infty \mathrm{d} k' \, k' \int_{k'}^\infty \frac{\mathrm{d} \chi'}{\chi'} \, K^2_{\mathrm{i}\tilde{\Omega}}\qty(\chi') \nonumber\\
			&\approx 0.000246.
		\end{align}
		Comparison of the approximation given by Eq.~\eqref{eq:jestem_wielki} and the exact result, given by Eq.~\eqref{eq:ours_expval_final}, is shown in Fig.~\ref{fig:part_vs_acc_massive}.
		We see that we reach the regime of validity of Eq.~\eqref{eq:jestem_wielki} almost immediately after satisfying the condition given in Eq.~\eqref{eq:sudden-birth}.

\section{Number of massless Unruh particles in a finite volume} 
\label{sec:number-massless}

	\begin{figure}[t]
		\centering
		\includegraphics[width=0.48\textwidth]{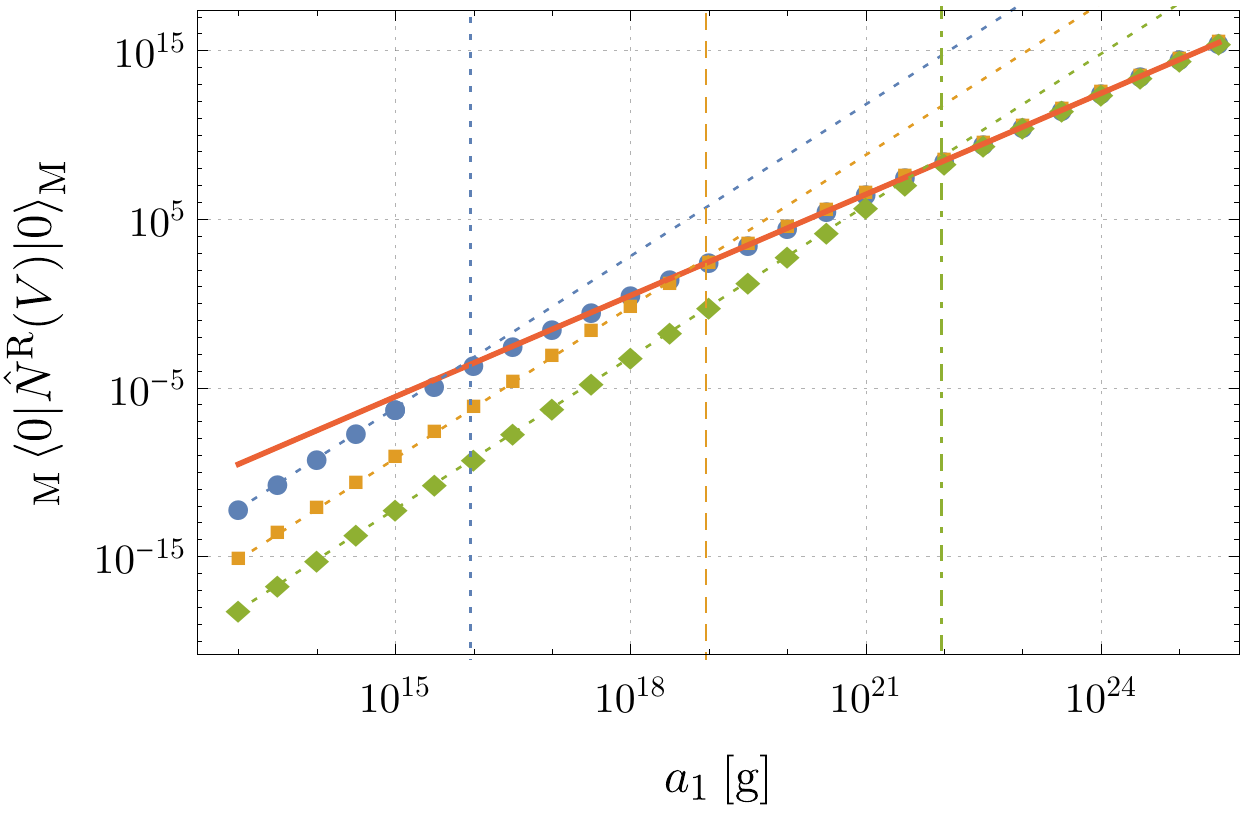}
		\caption{
			Average number of massless Unruh particles in a box of cross-section $1 \: \mathrm{m}^2$ as a function of its proper acceleration.
			The length of the box is $\chi_2 - \chi_1 = 1 \: \mathrm{m}$ (dots), $1 \: \mathrm{mm}$ (squares), and $1 \: \mathrm{\mu m}$ (diamonds).
			The continuous line is the high-acceleration limit given by Eq.~\eqref{eq:jestem_wielki}.
			The slanted dotted lines are the low-acceleration limits given by Eq.~\eqref{eq:massless_low_acc_limit}.
			The vertical lines denote smallest accelerations for which Eq.~\eqref{eq:validity_massless_low_acc} is satisfied, i.e., the length of the box becomes bigger than its distance from the Rindler horizon.
		}
		\label{fig:part_vs_acc_massless}
	\end{figure}

	As we have already remarked, the high acceleration limit given by Eq.~\eqref{eq:jestem_wielki} is mass-independent.
	In particular, it correctly describes the massless-field case, as depicted in Fig.~\ref{fig:part_vs_acc_massless}.
	Figure~\ref{fig:part_vs_acc_massless} also shows that the Unruh effect for massless particles has two distinct regimes: the $\big<\hat{N}^{\mathrm{R}}(V)\big> \propto a_1^2$ high-acceleration limit, and a $\big<\hat{N}^{\mathrm{R}}(V)\big> \propto a_1^3$ low-acceleration limit.
	To demonstrate how the latter follows from Eq.~\eqref{eq:ours_expval_final}, we use Eq.~\eqref{eq:final_chi'_k'} and expand it around $\chi_2/\chi_1 = 1$
	\begin{widetext}
	\begin{align} \label{eq:massless_low_acc_limit}
		_\mathrm{M}\langle 0| \hat{N}^\mathrm{R}(V) | 0\rangle_\mathrm{M} &\approx 0 + \qty(\frac{\chi_2}{\chi_1} - 1) \eval{\frac{\partial}{\partial \frac{\chi_2}{\chi_1}}}_{\frac{\chi_2}{\chi_1}=1} \frac{S_\perp}{2 \pi^3 \chi_1^2} \int \mathrm{d}\tilde{\Omega} \mathrm{d} k' \, k' \tilde{\Omega} \mathrm{e}^{-\pi \tilde{\Omega}} \int_{k'}^{k' \frac{\chi_2}{\chi_1}} \frac{\mathrm{d} \chi'}{\chi'} \,K^2_{\mathrm{i}\tilde{\Omega}}\qty(\chi') \nonumber\\ 
		&= \frac{V}{2 \pi^3 \chi_1^3} \int \mathrm{d}\tilde{\Omega} \mathrm{d} k' \, k' \tilde{\Omega} \mathrm{e}^{-\pi \tilde{\Omega}} K^2_{\mathrm{i}\tilde{\Omega}}\qty(k') \equiv D V a_1^3,
	\end{align}
	\end{widetext}
	where
	\begin{align} \label{eq:constant_D}
		D &\equiv \frac{1}{2 \pi^3} \int_0^\infty \mathrm{d}\tilde{\Omega} \, \mathrm{e}^{-\pi \tilde{\Omega}} \tilde{\Omega} \int_0^\infty \mathrm{d} k' \, k' K^2_{\mathrm{i}\tilde{\Omega}}\qty(k') \nonumber\\
		&\approx 0.000491.
	\end{align}
	The low acceleration limit, given by Eq.~\eqref{eq:massless_low_acc_limit} is shown in Fig.~\ref{fig:part_vs_acc_massless}.
	It turns out to be a good approximation whenever
	\begin{equation} \label{eq:validity_massless_low_acc}
	\chi_2/\chi_1 - 1 \leq 1,
	\end{equation}
	i.e., the length of the volume is smaller than its distance from the Rindler horizon.


\section{Discussion} 
\label{sec:discussion_outlook}

	The purpose of this article was to calculate how many particles a finite-size observer accelerating through the Minkowski vacuum would see.
	The motivation for this question is that any detector used to directly observe the Unruh effect will only count particles in a finite volume.
	To answer this question, one must employ a number operator that counts particles in a finite volume of space.
    Such an operator is not uniquely defined, but we have shown in Sec.~\ref{sec:operators} that two natural candidates yield equal expectation values for Fock states of the field, which includes the Minkowski vacuum.
	We used these operators in Secs.~\ref{sec:number-massive} and~\ref{sec:number-massless} to compute the number of particles in a finite volume as seen by an observer accelerating through the Minkowski vacuum (see Figs.~\ref{fig:part_vs_acc_massive} and~\ref{fig:part_vs_acc_massless}).

	For massive fields, we find a sudden birth of the Unruh effect occurs; that is, an appreciable number of particles only appears when the acceleration is higher than the threshold given by Eq.~\eqref{eq:sudden-birth}.
	This is equivalent to saying that the particles occur only in a tiny-layer, the thickness of which is of the order of the Compton wavelength, above the Rindler horizon.
{		For this reason the possibility of a direct observation of the Unruh effect for massive particles by an accelerated observer seems unlikely.}
	In this high acceleration regime, the number of particles was shown to be proportional to the square of the proper acceleration of the observer and to be independent of the mass associated with the field.

	For massless fields, two distinct regimes are observed: a high and a low acceleration regime.
	In the high acceleration regime, the average number of particles in a finite volume is the same as for a massive field.
	In the low acceleration regime, the average number of particles is proportional to the cube of the observer's acceleration.
	The transition to the high acceleration regime occurs when the distance from the Rindler horizon becomes smaller than the length of the volume, as given by Eq.~\eqref{eq:validity_massless_low_acc}.

\begin{acknowledgments}
The authors would like to thank Daniel R. Terno for useful discussions. This work was supported in part by the Natural Sciences and Engineering Research Council of Canada and the Dartmouth Society of Fellows. \\
\end{acknowledgments}
\appendix*

\section{Nonlocality of the particle density operator} 
\label{sec:nonlocality}

	In relativistic quantum field theory the density of particles is a nonlocal quantity~\cite{Bialynicki-BirulNJP2014,Mandel:1966,Redhead1995}.
	This means that the particle density operator given by Eq.~\eqref{eq:Min_Man_op} does not commute with itself for spacelike separations.
	Below we demonstrate that the same is true for the particle density defined by Eq.~\eqref{OurNumb}.
	We do this by working out the equal-time commutator of the density
	\begin{equation}
		\hat{\rho}(t,\bm{x}) \equiv  i \hat{\phi}^-(t, \bm{x}) \overset\leftrightarrow{\partial}_t \hat{\phi}^+(t, \bm{x})
	\end{equation}
	in Minkowski space.
	For brevity, in what follows we will write $\hat{\phi}^{\pm}_{1/2}$ and $\hat{\pi}^{\pm}_{1/2}$ instead of $\hat{\phi}^{\pm}(t,\bm{x}_{1/2})$ and $\hat{\pi}^{\pm}(t,\bm{x}_{1/2})$.

	We start by expanding $\comm{\hat{\rho}(t,\bm{x}_1)}{\hat{\rho}(t,\bm{x}_2)}$ using the bilinearity of the commutator and the Leibniz rule,
	\begin{widetext}
	\begin{align}
		\comm{\hat{\rho}(t,\bm{x}_1)}{\hat{\rho}(t,\bm{x}_2)} &= \comm{\hat{\phi}^{-}_1 \hat{\pi}^{+}_1}{\hat{\phi}^{-}_2 \hat{\pi}^{+}_2} + \comm{\hat{\pi}^{-}_1 \hat{\phi}^{+}_1}{\hat{\pi}^{-}_2 \hat{\phi}^{+}_2}  - \comm{\hat{\phi}^{-}_1 \hat{\pi}^{+}_1}{\hat{\pi}^{-}_2 \hat{\phi}^{+}_2} - \comm{\hat{\pi}^{-}_1 \hat{\phi}^{+}_1}{\hat{\phi}^{-}_2 \hat{\pi}^{+}_2} \nonumber\\
		&= \hat{\phi}^{-}_1 \comm{\hat{\pi}^{+}_1}{\hat{\phi}^{-}_2} \hat{\pi}^{+}_2 
		+ \hat{\phi}^{-}_2 \hat{\pi}^{+}_1 \comm{\hat{\phi}^{-}_1}{\hat{\pi}^{+}_2} 
		+ \hat{\pi}^{-}_1 \comm{\hat{\phi}^{+}_1}{\hat{\pi}^{-}_2} \hat{\phi}^{+}_2 
		+ \hat{\pi}^{-}_2 \hat{\phi}^{+}_1 \comm{\hat{\pi}^{-}_1}{\hat{\phi}^{+}_2} \nn \\
		&\quad  - \hat{\phi}^{-}_1 \comm{\hat{\pi}^{+}_1}{\hat{\pi}^{-}_2} \hat{\phi}^{+}_2 - \hat{\pi}^{-}_2 \hat{\pi}^{+}_1 \comm{\hat{\phi}^{-}_1}{\hat{\phi}^{+}_2} - \hat{\pi}^{-}_1 \comm{\hat{\phi}^{+}_1}{\hat{\phi}^{-}_2} \hat{\pi}^{+}_2 - \hat{\phi}^{-}_2 \hat{\phi}^{+}_1 \comm{\hat{\pi}^{-}_1}{\hat{\pi}^{+}_2} \nonumber\\
		&= \hat{\phi}^{-}_1 \comm{\hat{\phi}^{+}_2}{\hat{\pi}^{-}_1}^\dagger \hat{\pi}^{+}_2
		 - \hat{\phi}^{-}_2 \hat{\pi}^{+}_1 \comm{\hat{\phi}^{+}_1}{\hat{\pi}^{-}_2}^\dagger  + \hat{\pi}^{-}_1 \comm{\hat{\phi}^{+}_1}{\hat{\pi}^{-}_2} \hat{\phi}^{+}_2 
		 - \hat{\pi}^{-}_2 \hat{\phi}^{+}_1 \comm{\hat{\phi}^{+}_2}{\hat{\pi}^{-}_1}  \nn \\
		  &\quad 
		  - \hat{\phi}^{-}_1 \comm{\hat{\pi}^{+}_1}{\hat{\pi}^{-}_2} \hat{\phi}^{+}_2 + \hat{\pi}^{-}_2 \hat{\pi}^{+}_1 \comm{\hat{\phi}^{+}_2}{\hat{\phi}^{-}_1} - \hat{\pi}^{-}_1 \comm{\hat{\phi}^{+}_1}{\hat{\phi}^{-}_2} \hat{\pi}^{+}_2 + \hat{\phi}^{-}_2 \hat{\phi}^{+}_1 \comm{\hat{\pi}^{+}_2}{\hat{\pi}^{-}_1}. \label{eq:minkowski_comm_general}
	\end{align}
	\end{widetext}
	To calculate the remaining commutators, we use the field operators' modal decompositions given in Eq.~\eqref{posnegdec},
	\begin{align}
		\comm{\hat{\phi}^{+}_1}{\hat{\pi}^{-}_2} &= \int \mathrm{d}^3k \mathrm{d}^3k' \, u_{\bm{k}} (t,\bm{x}_1) \dot{u}^{\star}_{\bm{k}'} (t,\bm{x}_2) \comm{\hat{a}_{\bm{k}}}{\hat{a}^{\dagger}_{\bm{k}'}} \nonumber\\
		&= \int \mathrm{d}^3k \, \frac{1}{(2 \pi)^3 2 \omega_{\bm{k}}} \mathrm{i}\omega_{\bm{k}} \mathrm{e}^{\mathrm{i}\bm{k}(\bm{x}_{1}-\bm{x}_{2})} \nonumber\\
		&= \frac{\mathrm{i}}{2} \delta^{(3)}(\Delta\bm{x}),
	\end{align}
	where $\Delta\bm{x} \equiv \bm{x}_{1}-\bm{x}_{2}$.
	In a similar fashion, the commutator below gives
	\begin{align}
		\comm{\hat{\phi}^{+}_1}{\hat{\phi}^{-}_2} &= \int  \mathrm{d}^3k \mathrm{d}^3k' \, u_{\bm{k}} (t,\bm{x}_1) u^{\star}_{\bm{k}'} (t,\bm{x}_2) \nonumber\\
		&= \int \mathrm{d}^3k \, \frac{1}{(2 \pi)^3 2 \sqrt{\bm{k}^2+m^2}} \mathrm{e}^{\mathrm{i}\bm{k}\Delta\bm{x}}.
	\end{align}
	We now substitute $\bm{k}'=\bm{k}/m$, $\bm{x}'=m\bm{x}$ and calculate
	\begin{align}
		\comm{\hat{\phi}^{+}_1}{\hat{\phi}^{-}_2} &= \frac{m^2}{(2 \pi)^3 2} \int \mathrm{d}^3k' \, \frac{1}{\sqrt{\bm{k'}^2+1}}e^{i \bm{k'}\cdot\Delta\bm{x}'} \nonumber\\
		&= \frac{m^2}{(2 \pi)^3 2} \int \mathrm{d}\theta \mathrm{d}\varphi \mathrm{d}k' \, \frac{k'^2 \sin \theta}{\sqrt{k'^2+1}} \mathrm{e}^{\mathrm{i}k'\abs{\Delta\bm{x}'}\cos \theta} \nonumber\\
		&= \frac{m^2}{(2 \pi)^2 2} \int \mathrm{d}k' \, \frac{k'^2}{\sqrt{k'^2+1}} \int \mathrm{d}(\cos \theta) \, \mathrm{e}^{\mathrm{i}k'\abs{\Delta\bm{x}'}\cos \theta} \nonumber\\
		&= \frac{m^2}{2 \pi^2 \abs{\Delta\bm{x}'}} \int_0^\infty \mathrm{d}k' \, \frac{k'}{\sqrt{k'^2+1}} \sin k'\abs{\Delta\bm{x}'} \nonumber\\
		&= \frac{m^2}{2 \pi^2 \abs{\Delta\bm{x}}} K_1(m\abs{\Delta\bm{x}}) \label{eq:app_comm_fields},
	\end{align}
	where in the last equality we have used that~\cite{BatemanVol1}
	\begin{equation} \label{eq:bateman_sin}
		\int_0^\infty   \mathrm{d}x \, \frac{ x \sin(xy) }{ (x^2+\alpha^2)^{ \frac{3}{2} - \nu }} = \frac{\sqrt{\pi} (2 \alpha)^\nu  }{2\Gamma\left(\tfrac{3}{2}-\nu\right)}   y^{1 - \nu} K_\nu(\alpha y),
	\end{equation}
	which holds for $\Re\alpha,y>0$, $\Re \nu>-1$.
	Similarly, the commutator of momenta is
	\begin{align}
		\comm{\hat{\pi}^{+}_1}{\hat{\pi}^{-}_2} &= \int  \mathrm{d}^3k \, \dot{u}_{\bm{k}} (t,\bm{x}_1) \dot{u}^{\star}_{\bm{k}'} (t,\bm{x}_2) \nonumber\\
		&= \int \mathrm{d}^3k \, \frac{\sqrt{\bm{k}^2+m^2}}{(2 \pi)^3 2} \mathrm{e}^{\mathrm{i}\bm{k}\cdot \Delta\bm{x}} \nonumber\\
		&= \frac{m^3}{2 \pi^2 \abs{\Delta\bm{x}'}} \int_0^\infty \mathrm{d}k' \, k'\sqrt{k'^2+1} \sin k'\abs{\Delta\bm{x}'}, \label{eq:app_comm_momenta}
	\end{align}
	where in the last line we have used the same transformations as in Eq.~\eqref{eq:app_comm_fields}.
	Using Eq.~\eqref{eq:bateman_sin} now gives
	\begin{equation}
		\comm{\hat{\pi}^{+}_1}{\hat{\pi}^{-}_2} = - \frac{m}{2 \pi^2 \abs{\Delta\bm{x}}^2} K_2(m\abs{\Delta\bm{x}}).
	\end{equation}
	Inserting Eqs.~\eqref{eq:app_comm_fields} and~\eqref{eq:app_comm_momenta} into Eq.~\eqref{eq:minkowski_comm_general} gives
	\begin{align}
		\comm{\hat{\rho}(t,\bm{x}_1)}{\hat{\rho}(t,\bm{x}_2)} &= \frac{\mathrm{i}}{2} \delta^{(3)}(\Delta\bm{x}) \qty(\hat{\pi}^{-}_1 \hat{\phi}^{+}_2 - \hat{\phi}^{-}_1 \hat{\pi}^{+}_2) + \nonumber\\
		&\quad + \frac{\mathrm{i}}{2} \delta^{(3)}(\Delta\bm{x}) \qty(\hat{\phi}^{-}_2 \hat{\pi}^{+}_1 - \hat{\pi}^{-}_2 \hat{\phi}^{+}_1) + \nonumber\\
		&\quad + K_2(m\abs{\bm{x}}) \frac{m\qty(\hat{\phi}^{-}_1 \hat{\phi}^{+}_2 - \hat{\phi}^{-}_2 \hat{\phi}^{+}_1)}{2 \pi^2 \abs{\Delta\bm{x}}^2} + \nonumber\\
		&\quad - K_1(m\abs{\Delta\bm{x}}) \frac{m^2\qty(\hat{\pi}^{-}_1 \hat{\pi}^{+}_2 - \hat{\pi}^{-}_2 \hat{\pi}^{+}_1)}{2 \pi^2 \abs{\Delta\bm{x}}}.
	\end{align}
	We see that in addition to $\delta$-proportional terms, the commutator contains significant nonlocal contributions.
	This proves that the particle density defined by Eq.~\eqref{OurNumb} is nonlocal.

\bibliography{references} 

\end{document}